\documentclass[aps,nofootinbib,preprintnumbers,twocolumn,superscriptaddress,floatfix]{revtex4}%
\pdfoutput=1
\usepackage{amssymb}
\usepackage{epsfig}
\usepackage{bbm}
\usepackage{mathrsfs,graphicx,bm,amsmath}%
\usepackage{amsmath}%
\usepackage{amsfonts}%
\usepackage{graphicx}
\usepackage{bbm}
\usepackage{amssymb}
\usepackage{amsmath}
\usepackage{color}

\providecommand{\U}[1]{\protect\rule{.1in}{.1in}}

\newcommand{\I}{\mathrm{i}}
\newcommand{\be}{\begin{eqnarray}}
\newcommand{\ee}{\end{eqnarray}}

\newcommand{\nn}{\nonumber }

\begin{document}
\title{Finite-size and Particle-number Effects in an Ultracold Fermi Gas at Unitarity}

\author{Jens Braun}
\affiliation{\mbox{\it Theoretisch-Physikalisches Institut,
 Friedrich-Schiller-Universit{\"a}t Jena,}
\mbox{\it Max-Wien-Platz 1, D-07743 Jena, Germany}}

\author{Sebastian Diehl} 

\affiliation{Institut f\"ur Theoretische Physik, Universit\"at Innsbruck, A-6020 Innsbruck, Austria} 
\affiliation{Institute for Quantum Optics and Quantum Information 
of the Austrian Academy of Sciences, A-6020 Innsbruck, Austria} 

\author{Michael M. Scherer}

\affiliation{\mbox{\it 	Institut f\"ur Theoretische Festk\"orperphysik,
 RWTH Aachen,}
\mbox{\it Otto-Blumenthalstra\ss e, D-52074 Aachen, Germany}}

\begin{abstract}
We investigate an ultracold Fermi gas at unitarity confined in a periodic box~$V=L^3$ using renormalization group (RG) techniques.
Within this approach we can quantitatively assess the long range bosonic order parameter fluctuations which dominate finite-size 
effects. We determine the finite-size and particle-number dependence of universal quantities, such as the Bertsch
parameter and the fermion gap. Moreover, we analyze how these universal observables respond to the variation of an external pairing source. 
Our results indicate that the Bertsch parameter saturates rather quickly to its value in the thermodynamic limit as a function of increasing box size. 
On the other hand, we observe that the fermion gap shows a significantly stronger dependence on the box size, in particular for small values of 
the pairing source. Our results may contribute to a better understanding of  finite-size and particle-number effects  present in Monte-Carlo 
simulations of ultracold Fermi gases.
\end{abstract}
\maketitle

\section{Introduction}
Ultracold atoms provide accessible and controllable systems to study quantum many-body effects \cite{Bloch}. Often a single 
parameter, the s-wave scattering length~$a_{\rm s}$, describes the microscopic interactions completely, and can be tuned by 
means of an external magnetic field in the presence of a Feshbach resonance. In the context of ultracold fermions \cite{GiorginiRev}, 
this gives rise to the BCS-BEC crossover \cite{p5ALeggett80}, which interpolates between the cornerstones of quantum 
condensation phenomena, namely fermion superfluidity via Cooper pairing and Bose-Einstein condensation of molecular 
bosons, and has been probed in  milestone experiments \cite{Exp}.

The intermediate regime of the BEC-BCS crossover, where the scattering length is large and no obvious small 
expansion parameter exists, represents a major challenge for theoretical many-body approaches. Recently, experiments in this so-called 
unitary regime are reaching a level of precision \cite{PrecisionExp} that calls for an improved quantitative understanding 
on the theory side. 

Quantitative theoretical understanding can be obtained from numerical (Quantum) Monte Carlo 
simulations with the advantage of not relying on specific approximation schemes,
see e.~g. Refs.~\cite{Carlson:2003zz,Astrakharchik,Ceperley,Bulgac:2005pj,Burovski,Wingate:2006wy,Goulko:2009ny}.
On the other hand, various functional techniques exist which allow to associate 
specific physical mechanisms  with the resulting numbers for universal observables.
With these techniques, the complete phase diagram has been studied. In particular,
$\epsilon$-expansions~\cite{Nussinov, Nishida:2006br, Nishida:2006eu, EpsilonExpansion}, 
$1/N$-expansions~\cite{Nikolic:2007zz, AbukiBrauner}, $t$-matrix approaches~\cite{TMatrix}, Dyson-Schwinger equations~\cite{Diehl:2005ae, Diener08}, 
2-particle irreducible methods~\cite{Haussmann:2007zz}, and renormalization-group flow 
equations~\cite{Birse05, Diehl:2007th,%
Diehl:2007XXX, Gubbels:2008zz, Floerchinger:2008qc,Floerchinger:2009pg,Diehl:2009ma,KopLo09,Scherer:2010sv} have been employed.

Despite these tremendous theoretical efforts, our understanding of the limit of large scattering length is still not complete. 
In the following we shall focus on this limit for a spin-balanced two-component Fermi gas at zero temperature, and study the 
effects of both a finite volume and a finite particle number on universal quantities, such as the Bertsch parameter and the 
fermion gap. A key motivation for such an analysis is to foster the comparability of analytical approaches to data from 
lattice (Monte-Carlo) simulations which are performed in a finite volume. {In the context of resonantly interacting Fermi gases, 
finite-size and particle-number effects are currently under investigation using different Monte-Carlo
approaches~\cite{Wingate:2005xy,Wingate:2006wy,Goulko:2009ny,Forbes:2010gt,Lee:2010qp,Goulko:2010rw}. 
Of course, a comprehension of such effects may also provide useful information for an analysis of data from
Quantum Monte-Carlo studies~\cite{Drut:2010yn} of the Tan relations~\cite{Tan}. Let us also mention that
finite-volume effects in atomic few-body systems have been studied recently using an effective field theory 
approach~\cite{Kreuzer:2008bi,Kreuzer:2009jp,Bour:2011ef}. 
Finally, we would like to add that a study of finite-size and particle-number effects arising from a given experimental setup 
may eventually allow us to make better contact between theory and experiment, see e.~g. Refs.~\cite{Recati,Ku:2008vk}.}

For our analysis we employ a functional renormalization group (FRG) approach and apply techniques similar to those
used for the analysis of finite-volume effects in QCD low-energy models~\cite{Braun:2004yk,Braun:2005fj,Braun:2005gy,Braun:2010vd,Klein:2010tk}.
For our purposes, such an approach is advantageous since it allows us
to investigate the ``transition" between the finite system and its continuum limit. 

Within our RG approach, it is possible to continuously follow the change of relevant degrees of freedom by 
gradually integrating out the high momentum modes. Here, we are interested in quantifying long wavelength physics 
on the scale set by the finite size $L$, which typically exceeds the length scale set by the inverse Fermi momentum $k_\text{F}$, 
or by the chemical potential~$\mu$ defining the Fermi surface for positive values, 
$\sqrt{\mu} \lesssim k_\text{F}$, which is bounded from above by the Fermi 
momentum. The physics on these momentum scales well below the Fermi surface is dominated by bosonic fluctuations, which 
at zero temperatures physically are fluctuations of the superfluid order parameter in the crossover problem. 

We shall see below that our analysis fully supports this basic picture: Indeed, we find that the effects of a
finite system size are rather weak in a mean field approximation, which takes only fermion fluctuations into account but omits those 
of the order parameter. Taking the bosonic fluctuations into account, we find 
that the finite size effects on the Bertsch parameter are still rather weak, whereas the fermion gap is more sensitive. This 
complies with the general result that long wavelength phase fluctuations may have a strong impact on the condensate. In 
particular, as is well known, in lower spatial dimensions, these fluctuations may indeed
completely destroy the long range order.

From this argumentation, we see that our investigation can indeed help to quantitatively estimate the finite size errors of Quantum 
Monte Carlo simulations, which is notoriously difficult to improve on in lattice simulations. In particular, both the 
precise \emph{scaling} of an observable and the value of its \emph{relative deviation} for a given fixed size from the continuum 
results may be valuable in view of direct comparison, even if the absolute value of this observable in the continuum limit may not be
captured with high precision by our analytical approach due to an incomplete treatment of the short-range physics.

As an important technical development in our study,  we allow for a finite external (pairing) source~$J$ which couples to the order-parameter field. 
The relevance of such an external source for studies in a finite volume has been pointed out in Refs.~\cite{Chen:2003vy,Wingate:2005xy}.
In any case, the inclusion of a pairing source allows us to control symmetry breaking in a finite volume.
As done in lattice (Monte-Carlo) simulations, see e.~g. Ref.~\cite{Bulgac:2005pj}, we shall choose
the boundary conditions of the fermions in the spatial directions to be periodic.  
Moreover, we consider the grand canonical ensemble for our studies. This implies that our choice for the chemical potential fixes 
the (average) particle number of the system.

Finally, we add that we do not study possible finite-density corrections in lattice simulations which arise from
the discretization of space-time and are associated with modifications of the (continuum) dispersion relation on the lattice.
This type of finite-size effects has been addressed in a recent Dynamical Mean Field (DMFT) study~\cite{Privitera10}.

{The paper is organized as follows: In Sect.~\ref{sec:gen}, we discuss general aspects of finite-size effects in
ultracold Fermi gases at unitarity. A mean-field analysis of finite-size and particle-number effects is then presented 
in Sect.~\ref{sec:mf}.  In Sect.~\ref{sec:bmf} we study the impact of order-parameter
fluctuations in a finite system, i.~e. we take into account effects arising from the Nambu-Goldstone modes in the spectrum
of the theory. Our conclusions and a brief outlook are given in Sect.~\ref{sec:conc}.

\section{General Aspects of effects of a finite volume and a finite external source}\label{sec:gen}

Let us start our general discussion of finite-size and particle-number effects in a Fermi gas at unitarity by considering
a standard mean-field approximation.} We emphasize, however, that the general finite-size formalism developed here is directly 
applicable to the analysis beyond mean field in the Sect.~\ref{sec:bmf}. 
The mean-field approximation amounts to dropping 
contributions associated with fluctuations of the order-parameter field.
In Gross-Neveu-type models, for example, such an approximation
corresponds to the leading order contribution in a systematic expansion in powers of~$1/N_{\rm f}$, 
where $N_{\rm f}$ denotes the number of fermion species~\cite{Braun:2010tt}. 
Such an approximation already allows us to discuss finite-size effects as far as they relate to the fermionic modes
of the system. The role of the dynamics of bosonic bound-states of fermions
will be discussed quantitatively in Sect.~\ref{sec:bmf}.

The microscopic (classical) action~$S$ of a resonantly interacting Fermi gas can be written as follows:
\be
&& S [\psi^{\dagger},\psi]=  \int d\tau\int d^3 x\, \left\{ \psi^{\dagger}
\left( \partial _{\tau} - \vec{\nabla}^{\,2} -\mu \right) \psi\right. \nn\\
&&\qquad\qquad\qquad\qquad\qquad\quad
 \left. +\,\frac{1}{2}\bar{\lambda}_{\psi}(\psi^{\dagger}\psi) (\psi^{\dagger}\psi)
\right\}\,,\label{eq:Faction}
\ee
where $\psi^{\rm T}=(\psi_{\uparrow},\psi_{\downarrow})$ and $\bar{\lambda}_{\psi}$ denotes the bare 
four-fermion coupling. The chemical potential is given by~$\mu$.
Note that~$2m=1$ in our conventions, where~$m$ is the mass of the fermions.

The dimensionless renormalized four-fermion coupling~$\lambda_{\psi}\sim \bar{\lambda}_{\psi}k$ in Eq.~\eqref{eq:Faction} 
is related to the s-wave scattering length~$a_{\rm s}$, where~$k$ is the RG scale. To be specific, we have
\be
\lambda_{\psi}=  \frac{8\pi\Lambda}{\frac{1}{a_{\rm s}}-c_{\rm reg.}\Lambda}
\, \label{eq:lpsiSL}
\ee
in the limit of a broad Feshbach resonance,
where~$\Lambda$ denotes the ultraviolet (UV) cutoff. The constant~$c_{\rm reg.}>0$ depends on the employed 
regularization scheme. For example, we have~$c_{\rm reg.}=2/\pi$ for the sharp UV cutoff.

In order to study the RG flow of the quantum effective action~$\Gamma$, we employ a non-perturbative RG equation, the 
Wetterich equation~\cite{Wetterich:1992yh}. The effective action then depends on the RG scale~$k$ (infrared cutoff scale) which
determines the RG ``time" $t=\ln(k/\Lambda)$. For reviews on and introductions to this functional RG approach we refer the 
reader to Refs.~\cite{Litim:1998nf,Bagnuls:2000ae,Berges:2000ew,Polonyi:2001se,Delamotte:2003dw,%
Pawlowski:2005xe,Gies:2006wv,Delamotte:2007pf,Diehl:2009ma,Rosten:2010vm,Honerkamp5,Braun:2011pp}.

For our analysis of finite-size and particle-number effects, we employ 
the following {\it ansatz} for the effective action:
\be
&& \!\!\!\!\!\!\Gamma_k [\psi^{\dagger},\psi,\varphi]\! =\! \int d\tau\! \!\int d^3 x\, \Big\{ \psi^{\dagger}
\left( \partial _{\tau} \! -\!  \vec{\nabla}^{\,2}\! -\!\mu \right) \psi 
\!+\!  \bar{m}_{\varphi}^2 \varphi^{\ast}\varphi\nn\\
&& \qquad\qquad\qquad\qquad\quad
+ \frac{1}{2}\bar{\lambda}_{\varphi}(\varphi^{\ast}\varphi)^2
+ \frac{1}{\sqrt{2}}J(\varphi+\varphi^{\ast})
\nn\\
&& \qquad\qquad\qquad\qquad\quad\;
- \bar{h}_{\varphi}\left[ \varphi^{\ast}\psi_{\uparrow}\psi_{\downarrow}\! -\! \varphi\, \psi_{\uparrow}^{\ast}\psi_{\downarrow}^{\ast}\right]
\Big\}\,,\label{eq:Gmf}
\ee
where~$J$ denotes an external pairing source and the coefficient of the cubic fermion-boson interconversion term, $\bar{h}_{\varphi}$, 
is a Feshbach or Yukawa coupling. Note that the source term~$\propto J$ 
explicitly breaks the underlying~U($1$) symmetry of the microscopic theory.
Formally, this ansatz is inspired by the partially bosonized formulation of the action~\eqref{eq:Faction}, which 
can be obtained by introducing a complex scalar field~$\varphi \sim (\bar{h}_{\varphi}/\bar{m}^2_{\varphi})(\psi_{\uparrow}\psi_{\downarrow})$ 
into the path integral by means of a Hubbard-Stratonovich transformation. Physically, this model describes the physics of fermionic 
atoms close to a Feshbach resonance in an intuitive way, where the $\bar m_\varphi^2$ plays the role of the detuning from a bosonic 
bound state level, described by the complex field~$\varphi$, at the microscopic scale. 
The bare couplings $\bar{h}_{\varphi}$ and~$\bar{m}_{\varphi}$ are {\it a priori}
constants at our disposal and are chosen such that the four-fermion interaction term in Eq.~\eqref{eq:Faction} 
vanishes identically,~$\bar{\lambda}_{\psi}=-\bar{h}_{\varphi}^2/\bar{m}_{\varphi}^2$. The model becomes physically equivalent to 
the model (\ref{eq:Faction}) in the broad resonance limit, where $\bar h_\varphi\to \infty$, while the ratio $\bar{h}_{\varphi}^2/\bar{m}_{\varphi}^2$ is 
kept fixed \cite{Diehl:2005ae} on the microscopic scale, which we consider in this work.  
For our purposes below, it is convenient to consider two real-valued fields~$\varphi_1$ and~$\varphi_2$ instead of
the two complex-valued fields~$\varphi$ and~$\varphi^{\ast}$, where $\varphi=(\varphi_1+\I\varphi_2)/\sqrt{2}$.
We would like to add that a partially bosonized formulation is beneficial since it allows us to 
resolve (parts of) the momentum dependence of fermionic interactions by means of a straightforwardly accessible 
derivative expansion of the effective action. 

Due to quantum fluctuations, the cubic interaction generates kinetic terms for the bosonic fields in the RG flow,
\be
Z_{\varphi}^{\|}\varphi^{\ast}\partial_{\tau}\varphi
\quad\text{and}\quad
Z_{\varphi}^{\perp}\varphi^{\ast}\vec{\nabla}^{\,2}\varphi\,, 
\ee
even if these terms have been set to zero at the initial RG scale~$\Lambda$. For example, taking these fluctuations into 
account by following the RG flow down to $k\to0$, and concentrating on the BEC regime of the crossover for a moment, 
where the bosonic field describes tightly bound molecular states, their ratio 
is given by $Z_{\varphi}^{\perp}/Z_{\varphi}^{\|}\to 1/2$ or, with dimensions restored, $1/(4m)$. Indeed, this signals propagating 
bosonic bound states of double mass of the fermionic atoms. More generally, the leading momentum-dependence of the four-fermion 
vertex  in Eq.~\eqref{eq:Faction} in the pairing channel is encoded in these kinetic terms. In leading order in a systematic 
expansion of~$\Gamma$ in derivatives, 
the RG flows of the wave-function renormalizations~$Z_{\varphi}^{\|,\perp}$
are essentially determined by a purely fermionic one-particle irreducible (1PI) diagram in the U($1$) symmetric regime. Thus, we have
\be
\partial _t \ln Z_{\varphi}^{\|,\perp} = -c_{\varphi} \bar{h}_{\varphi}^2\,,\label{eq:eta}
\ee
where $c_{\varphi}$ is a positive constant~\cite{Diehl:2009ma}. 
Contributions from 1PI diagrams with internal boson lines are suppressed 
in the symmetric regime due to the large (renormalized) boson mass 
parameter~$m_{\varphi}=\bar{m}_{\varphi}^2/(Z_{\varphi}^{\perp}k^2)$, at least in the limit of a broad 
Feshbach resonance. For a similar reason, the running of the fermionic wave-function renormalizations is subleading
since corrections due to 1PI diagrams with at least one internal boson and fermion line are
suppressed, both in the U($1$) symmetric regime as well as in the regime with broken U($1$) 
symmetry in the ground state. Finally, the RG flow equation of the Yukawa coupling also assumes a simple form
in the symmetric regime, since it is only driven by the anomalous dimensions of the bosonic 
fields:
\be
\partial_t h^2_{\varphi} = (\eta^{\perp}_{\varphi}-1)h^2_{\varphi}\,,
\ee
where $h_{\varphi}^2=\bar{h}_{\varphi}^2/(Z_{\varphi}^{\perp}k)$, see Ref.~\cite{Diehl:2009ma}. 
Note that we are free to choose either~$Z_{\varphi}^{\|}$ or~$Z_{\varphi}^{\perp}$ to renormalize the Yukawa coupling. 
In the present work, we only take into account the running of~$Z_{\varphi}^{\|}$ and 
set~$Z_{\varphi}^{\|}=Z_{\varphi}^{\perp}$. We would like to add that the values of low-energy observables,
such as the fermion gap and the Bertsch parameter, do not depend on the RG flow of the 
wave-function renormalizations in the mean-field limit. Beyond this limit, however, the running of the latter affect the
running of physical observables.

The universal behavior associated with the limit of a large (infinite) s-wave scattering length~$a_{\rm s}$
is linked to the existence of a non-trivial UV attractive fixed-point of the theory~\cite{Nikolic:2007zz,Diehl:2007XXX}. In a given
regularization scheme, the value of the four-fermion coupling given in Eq.~\eqref{eq:lpsiSL} can be associated with
the ratio of the UV fixed-point values~$\bar{h}_{\varphi}^{\ast}$ and~$\bar{m}_{\varphi}^{\ast}$ of
the Yukawa coupling and the mass parameter, respectively. Recall that~$\bar{\lambda}_{\psi}=-\bar{h}_{\varphi}^2/\bar{m}_{\varphi}^2$.
In order to study the limit of infinite s-wave scattering length,  we therefore have to choose 
the initial conditions for  the RG flow equations of~$\bar{m}_{\varphi}$ and~$\bar{h}_{\varphi}$ 
such that they are close to their UV fixed point values, see e.~g. Refs.~\cite{Diehl:2007XXX,Diehl:2009ma,Braun:2011pp}.

In our ansatz~\eqref{eq:Gmf} we also allow for a term~$\propto \bar{\lambda}_{\varphi}$ which 
describes four-boson interactions. At the initial
RG scale~$\Lambda$ we set the coupling~$\bar{\lambda}_{\varphi}$ to zero. For scales~$k<\Lambda$, this coupling is then generated due to 
quantum corrections, even if we do {\it not} take into account fluctuations of the bosonic fields. Boson self-interactions
of higher order~$(\varphi^{\ast}\varphi)^m$ with $m>2$ are also generated. We add that the bosonic self-interaction terms (with~$m\geq 1$) 
essentially parameterize the effective potential (order-parameter potential)~$U$. In the following we drop the
contributions from higher-order interactions ($m>2$) for simplicity. We rush to add that 
the RG flows of the associated couplings are decoupled in the mean-field limit
in the U($1$) symmetric regime at scales~$k>k_{\rm SB}$, where~$k_{\rm SB}$ denotes the scale associated with spontaneous
U($1$) symmetry breaking.\footnote{For a comprehensive discussion we refer the reader to Ref.~\cite{Braun:2010tt}
where this issue has been studied in detail in the Gross-Neveu model. Since these arguments only rely on very general 
properties of partially bosonized formulations of fermionic theories, the same reasoning also holds in the present study.}
Thus, the determination of the scale~$k_{\rm SB}$
does not suffer from this approximation, at least in the mean-field limit.
However, the (exact) value of a given low-energy observable~$\mathcal O\sim k_{\rm SB}^2$
may receive contributions from boson self-interactions of higher order.\footnote{Here,
we have tacitly assumed that~${\mathcal O}$ has the dimension of energy.}
As we shall see below, these  corrections to the fermion gap and the Bertsch parameter 
are small. Again, this holds at least in the mean-field approximation.
 
As already indicated above, we also include a source~$J$ for the field~$\varphi_1$ in our RG study. This term breaks explicitly
the underlying~U($1$) symmetry of the theory. However, such a linear symmetry 
breaking term remains unchanged in the RG flow \cite{Zinn-Justin:2002ru}. Instead of including such a term in the RG
flow, one may therefore simply study the evolution of the effective action without a finite external source~$J$. Explicit symmetry breaking
can then be taken into account after the quantum fluctuations have been integrated out on all scales. This strategy has been 
followed in the context of low-energy models of QCD, see e.~g. Refs.~\cite{Berges:1997eu,Schaefer:1999em}. 
In particular, such an approach is perfectly suited when one has access to the 
full effective potential~$U(\bar{\rho})=T\Gamma[\bar{\rho}]/L^3$ including
bosonic self-interactions of arbitrarily high orders, where~$\bar{\rho}=(\bar{\varphi}^{\ast}\bar{\varphi})$;
$\bar{\varphi}$ denotes a spatially constant background field and~$1/T$ is the extent of the system
in Euclidean time direction. Using a low-order expansion of the effective potential as indicated in Eq.~\eqref{eq:Gmf}, such an approach still yields 
reasonable results in an infinite volume and for small values of the source~$J$.
In a finite-volume study, however, the situation is different.
Without a finite value for~$J$, fluctuations of massless Nambu-Goldstone bosons associated 
with spontaneous U($1$) symmetry breaking restore the symmetry in the (deep) IR limit.
This implies that there is no spontaneous~U($1$) symmetry breaking in a finite volume. 
Therefore we include a finite source~$J$
on all scales in the RG flow to control~U($1$) symmetry breaking in a finite system.\footnote{Due to our
definition of~$J$, we have~$\langle \phi \rangle_{J}=\langle \phi_1\rangle_{J}/\sqrt{2}>0$ 
and~$\langle \phi_2\rangle=0$ for~$J < 0$.}
This renders the order-parameter~$\langle \phi\rangle_{J}\sim\langle\phi_1\rangle_{J}$
finite on all scales.\footnote{In this work, we employ the convention 
that the field~$\phi$ used in the expectation value~$\langle\, \dots \rangle$ denotes the quantum field which appears in the
path integral. This quantum field should not be confused with the so-called classical field~$\varphi$ which appears in the effective action.}

Let us briefly discuss a subtlety concerning the expectation values~$\langle \phi \rangle_{J}$ 
and~$\langle|\phi|\rangle\equiv \langle |\phi|\rangle_{J=0}$. In the limit~$J\to 0$, the 
minima of the order-parameter potential~$U$ are degenerate and describe a
circle of radius~$\langle |\phi|\rangle$ in the plane spanned by~$\bar{\varphi}_1$ and~$\bar{\varphi}_2$. 
For a finite value of~$J$, the degeneracy of the ground states is completely lifted and the mass $m_{\rm G}$ of the
pseudo Nambu-Goldstone modes is finite.
We will now argue that~$|\langle \phi \rangle|:=|\lim_{J\to 0}\langle \phi \rangle_{J}|$
and $\langle|\phi|\rangle$ agree in the infinite-volume limit. In this limit,
both quantities may therefore be considered as equivalent order parameters for spontaneous U($1$) symmetry breaking.
In order to understand better the relation between these two quantities, let us first consider a system in a finite
periodic volume~$V$. In this case, loosely speaking, one finds
that the fluctuations of the spatial (momentum) zero-modes of the Nambu-Goldstone fields
along the circle of energetically equivalent ground states
tend to restore the symmetry. Schematically, these fluctuations are suppressed according 
to\footnote{For the Goldstone bosons with linear dispersion, the relativistic 
power counting, where $m_{\rm G}$ corresponds to an inverse length, is appropriate.}
$\sim\exp(-(m_{\rm G}\sqrt[3]{V})^{c})$, where~$m_{\rm G}$ denotes the mass of the (pseudo) Nambu-Goldstone particles for~$J\neq 0$
and $c\geq 1$ is a positive constant~\cite{Gasser:1987ah}. 
Considering the limit~$m_{\rm G}\to 0$ ($J\to 0$) before
taking the limit~$V\to\infty$, we then anticipate that $\langle\phi\rangle_{J=0}$ 
averages to zero since the fluctuations of the zero modes are not suppressed. On the other
hand, these fluctuations of the zero modes are projected out in the computation of~$\langle |\phi|\rangle$.
In a finite volume at sufficiently low temperatures, $\langle|\phi|\rangle$ can therefore be finite for $J=0$, even 
if we have~$\langle\phi\rangle_{J=0}\!=\! 0$. In fact, we always have $\langle \phi\rangle_{J=0}=0$ in a finite volume. 
Considering now the limit~$V\to\infty$ and then~$m_{\rm G}\to 0$, fluctuations of the 
Nambu-Goldstone along the circle of energetically equivalent ground states 
are exponentially suppressed and $\langle \phi\rangle_{J=0}$
remains finite. In particular, we have $|\langle \phi \rangle|=\langle|\phi|\rangle$. 
For sufficiently large volumes, however, it is reasonable to assume that the deviations of $\langle |\phi|\rangle$
from the value of $|\langle \phi\rangle|$ are small in the thermodynamic limit.\footnote{Note that fluctuations of the
Nambu-Goldstone are not included in a mean-field study. This implies that~$\langle \phi\rangle_{J}$ remains finite in a
finite volume at sufficiently low temperatures, even for $J\to 0$.}
In the following we always use~$\langle\phi\rangle_{J}$ for $J\neq 0$ 
as a (pseudo) order parameter for U($1$) symmetry breaking in a finite volume.
Note that the fermion gap and the ground-state energy of the Fermi gas depend on the actual value of~$\langle \phi\rangle_{J}$.

{Finally a comment on} inhomogeneous ground states is in order. In a finite volume and for a finite particle number 
the ground state of the theory might be inhomogeneous, implying that~$\langle \phi\rangle_{J}$ has a 
non-trivial dependence on the spatial coordinates.\footnote{The ground state of a resonantly interacting Fermi gas
may be inhomogeneous even in the infinite-volume limit. For example, a Sarma-type~\cite{Sarma} or Fulde-Ferrell-Larkin-Ovchinnikov~\cite{Fulde} phase
may exist for unequal chemical potentials of the spin-up and spin-down fermions.} 
From studies of relativistic theories, 
there is indeed direct evidence that finite-size effects may alter the phase structure of a given theory.
For example, lattice studies of the $1+1$d Gross-Neveu model show that the finite-temperature
phase diagram of the uniform system is modified significantly due to the non-commensurability of 
the spatial lattice size with intrinsic length scale of the inhomogeneous condensate~\cite{deForcrand:2006zz}.
However, a study of the emergence of an inhomogeneous ground state as a 
function of~$N$ and~$L$ is beyond the scope of the present work. From here on, 
we shall assume that the ground state of the theory is homogeneous, even for finite values of~$N$ and~$L$. Thus, we consider
the (pseudo) order-parameter~$\langle \phi\rangle_{J}$ to be independent of the space-like coordinates.
This amounts to exploring only
leading-order effects on physical observables arising from the presence of a finite volume~$V$ and a finite particle number~$N$. At least close
to the thermodynamic limit, it still seems reasonable to expect that corrections due to possible inhomogeneities of the ground state are subleading.

\section{Mean-Field Analysis of Finite-Size and Particle-Number Effects}\label{sec:mf}

Let us now discuss our RG study in the mean-field limit in more detail.
The flow equation of the order-parameter potential 
in the mean-field limit can be derived along the lines of Ref.~\cite{Diehl:2009ma}, where the thermodynamic limit 
has been studied in detail. For a system in a finite volume we then obtain:
\be
\partial_t U(\bar{\rho},J,L,\mu)=-2k^5 (B^{>}_{\rm F}+B^{<}_{\rm F})s_{\mathrm{F}}\,,
\label{eq:mfflowU}
\ee
where the functions~$B_{\rm F}^{>}$~and~$B_{\rm F}^{<}$ are given by
\be
B_{\rm F}^{>} &=& \frac{1}{(kL)^3}\sum_{\vec{q}}\theta \left((kL)^2-(2\pi)^2\vec{q}^{\,2}+\mu L^2\right)\nn\\
&&\qquad\qquad\qquad\qquad\;\times \theta \left((2\pi)^2\vec{q}^{\,2}-\mu L^2\right)\,,\label{eq:BFg}\\
B_{\rm F}^{<} &=& \frac{1}{(kL)^3}\sum_{\vec{q}}\theta \left((kL)^2+ (2\pi)^2\vec{q}^{\,2}-\mu L^2\right)\nn\\
&&\qquad\qquad\qquad\qquad\; \times\theta \left(\mu L^2 - (2\pi)^2\vec{q}^{\,2}\right)\,.\label{eq:BFs}
\ee
These functions count the momentum modes. Recall that we use periodic boundary conditions
for the fermions in spatial directions. Furthermore, we have to specify the function $s_{\mathrm{F}}$ at $T=0$,
\be
s_{\mathrm{F}}=\frac{k^2}{\sqrt{k^4+\bar h_{\varphi}^2\bar\rho}}\,.
\ee
The fermion gap is given by~$\Delta= \bar h_{\varphi}^2\bar\rho_0=h_{\varphi}^2|\langle\phi\rangle_{J}|^2$.
To regularize the theory, we use here and
in Sect.~\ref{sec:bmf} an optimized regulator function~\cite{Diehl:2009ma}.
For details on optimization of RG flows, we refer the reader to 
Refs.~\cite{Litim:2000ci,Litim:2001fd,Litim:2001up,Pawlowski:2005xe}.

A quantity of utmost importance in our study is the fermion density~$n$.
The RG flow of the density~$n$ can be deduced from the flow equation of the effective potential. For details, 
we refer the reader to Ref.~\cite{Diehl:2009ma}. The (average) particle number~$N$ is then given by~$N=nL^3$.
The initial condition for the RG flow of the density is given by the  
one of the free Fermi gas~$n_{\text{free}}$, which is determined by our  
choice for the chemical potential. To be specific,~$n_{\text{free}}$ is given by
\be
n_{\text{free}}=\frac{s}{L^3}\sum_{\vec{q}}\theta \left( \mu L^2 \!-\!  
(2\pi)^2 \vec{q}^{\,2}
\right)\;\stackrel{(\mu L^2\to\infty)}{\longrightarrow}
\;\frac{\mu^{\frac{3}{2}}}{3\pi^2}
\,,\label{eq:nfree}
\ee
where~$q_i \in \mathbb{Z}$ and $s$ is the spin-degeneracy factor. For  
a two-component Fermi gas, we have~$s=2$. Apparently, $n_{\text{free}}$ is discontinuous for 
finite values of~$L$. We shall come back to this below.

For clarity, let us discuss the definition of the Bertsch parameter 
and the (dimensionless) fermion gap which are at the heart of our interest in the present work.
In the limit of large s-wave scattering lengths~$a_{\rm s}$, (dimensionless) IR observables are universal,
i.~e. they do not depend on the actual value of the density~$n$, provided that we consider~$J\to 0$ 
in the thermodynamic limit. In our numerical studies,~$\mu$ represents a dimensionful parameter which can be adjusted by hand.
For a given value of~$\mu$, we then compute the density~$n$ of the interacting system.
This density can then be related to the Fermi energy~$\epsilon_{\rm F}$ of a free Fermi 
gas with the same density:~$\epsilon_{\rm F}=k_{\rm F}^2$, where the Fermi 
momentum is given by~$k_{\rm F}=(3\pi^2 n)^{1/3}$. Since we consider the limit~$a_{\rm s}\to\infty$, the energy~$E$ of the 
interacting system is proportional to~$\mu$, as it is the case for the free system. Thus, we have~$E/N= \xi E_{\rm F}/N$, 
where~$\xi$ is the so-called Bertsch parameter. For the free system,
we have~$E_{\rm F}=\epsilon_{\rm F}^{5/2}V/(5\pi^2)$. The energy per particle is given by~$E_{\rm F}/N=(3/5)\epsilon_{\rm F}$. 
Since~$\mu\neq \epsilon_{\rm F}$ for the interacting system, it follows that 
\be
\frac{E}{N}=\frac{3}{5}\xi\epsilon_{\rm F}\quad\text{and}\quad \mu = \xi \epsilon_{\rm F}=\xi (3\pi^2 n)^{\frac{2}{3}}\,,
\label{eq:xidef}
\ee
where~$E/N:=(3/5)\mu$.
The dimensionless fermion gap~$\bar{\Delta}$~can be defined accordingly by using the Fermi energy of the free system as a 
reference scale,~$\bar{\Delta}=\Delta/\epsilon_{\rm F}$. In our studies of a Fermi gas in a finite volume~$V\!=\!L^3$ in the presence of a finite 
pairing source~$J$, the energy~$E$ depends on three scales, namely~$\mu$, $L$ and~$J$. For a given set of values
for~$\mu$,~$L$ and~$J$, we can then compute the density~$n=n(\mu,L,J)$ and the average particle number~$N=nL^3$ 
of the interacting system. In order to ``measure" the energy of the 
interacting Fermi gas, we again use the Fermi energy~$\epsilon_{\rm F}$ of a free uniform Fermi gas as a reference scale. 
For convenience, we {\it define} the energy per particle of the interacting system as~$E/N:=(3/5)\mu(N,L,J)$.
In analogy to the thermodynamic limit, we can then define the Bertsch parameter as follows
\be
\frac{E(N,L,J)}{N}&=& \frac{3}{5}\xi(N,L,J) \epsilon_{\rm F}\quad\text{and}\nn\\
\mu(N,L,J) &=& \xi (N,L,J) (3\pi^2 n)^{\frac{2}{3}}\,,\label{eq:xidef2}
\ee
where~$n$ is the density of the interacting finite system. From~$\xi (N,J,L)$ we can then read off how
the energy (density) of the free uniform system changes when we vary~$N$,~$J$ and~$L$.
The dimensionless gap in a finite volume is again
defined by using the Fermi energy~$\epsilon_{\rm F}=(3\pi^2 n)^{2/3}$ of a free uniform gas as a reference scale, 
where the density of this free gas is identical to the density of the interacting finite system.
Our observations regarding the volume dependence of the Bertsch parameter and the fermion gap
can in principle be used to guide Monte-Carlo simulations of the type 
discussed in Ref.~\cite{Chen:2003vy}. However, our observations may also provide useful insights 
into the~$N$- and~$L$-dependence of results from other Monte-Carlo approaches.
\begin{figure}
\includegraphics[width=1\linewidth]{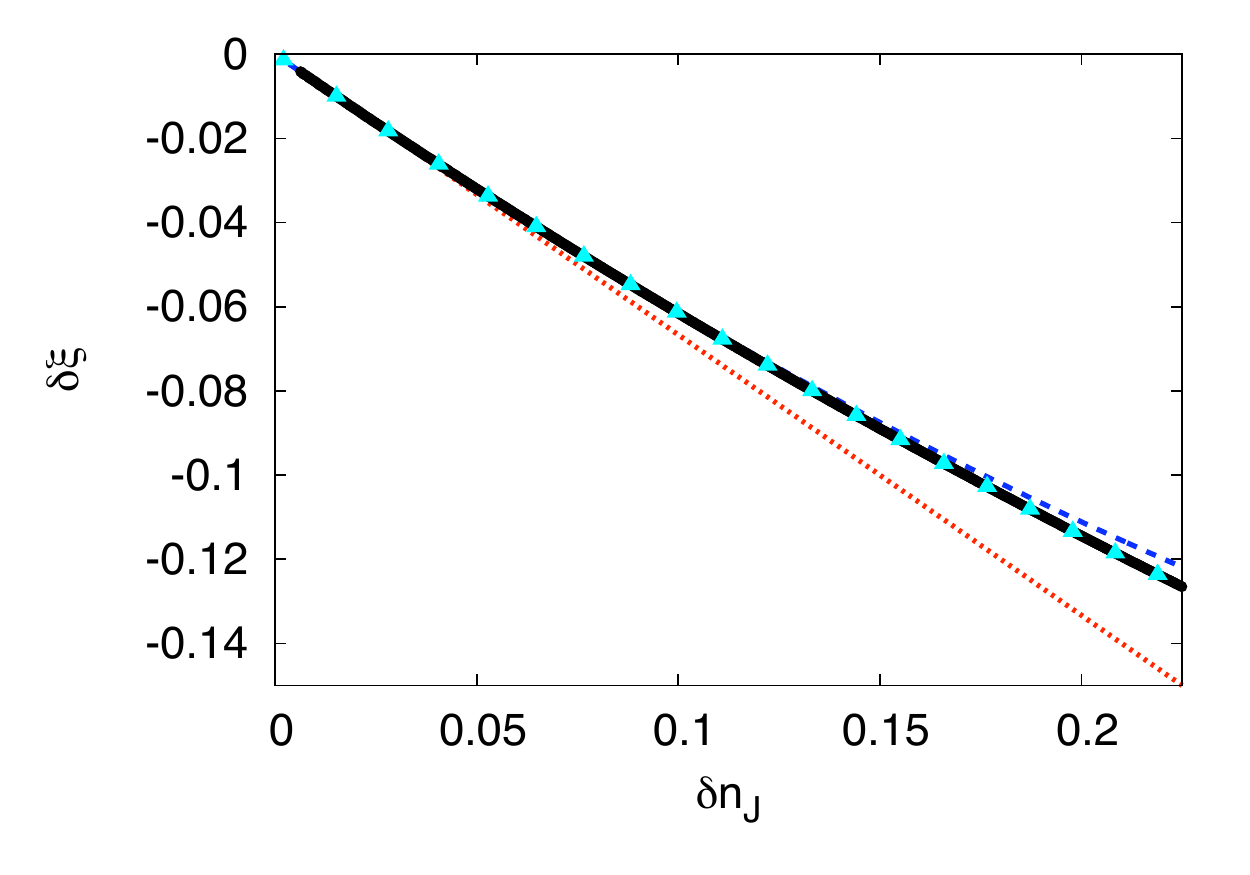}
\caption{\label{fig:dbdn} Relative shift of the Bertsch parameter~$\delta\xi$ 
as a function of~$\delta n_J=(n_J-n_{J=0})/n_{J=0}$ in the thermodynamic limit. The black solid line
depicts the numerical results from our mean-field study 
for~$J k_{\rm F}^{-7/2}=0.007\dots 0.7$~(or equivalently~$J\mu^{-7/4}=0.003\dots 0.3$).
The triangles are the results for~$\delta n_J$ from our study including order-parameter fluctuations, 
where~$J k_{\rm F}^{7/2}=0.009\dots 1.023$, see Sect.~\ref{sec:bmf}. 
While the absolute values for~$\xi$ differ from
the ones of the mean-field study, the results for~$\delta \xi$ are indistinguishable from 
the mean-field results on the scale of the plot.
The red dotted line and the blue dashed line represent
the results from the expansion of~$\delta\xi$ up to first and second order in~$\delta n_J$, 
respectively, see Eq.~\eqref{eq:deltaxi}.
}
\end{figure}

Before we study the Fermi gas in a finite volume, we briefly discuss the effect of a finite
(pairing) source~$J$ in the thermodynamic limit. 
To this end, we analyze the $J$-dependence of the fermion gap and the Bertsch parameter.
While the Bertsch parameter can be viewed as a thermodynamic observable, which
depends only indirectly on~$J$, the fermion gap plays the role of the order parameter for~U($1$)
symmetry breaking and is therefore directly related to the pairing source~$J$. 
For~$J\to 0$, we obtain~$\xi\approx 0.60$ and~$\Delta/\epsilon_{\rm F}\approx 0.63$
within our mean-field approximation.\footnote{Our result for the fermion gap does not agree with
the ``exact" mean-field result for this quantity, see e.~g. Refs.~\cite{p5ALeggett80,KopLo09,2010arXiv1008.5086H}, 
since we have dropped contributions from bosonic self interactions~$(\varphi^{\ast}\varphi)^{m}$ of higher order ($m>2$), 
see our discussion above. In the regime with broken~U($1$) symmetry, 
the RG flows of bosonic self-interactions~$(\varphi^{\ast}\varphi)^{m}$ are coupled
even in the mean-field limit due to the presence of a finite vacuum expectation value~$\langle \phi\rangle$.
Note that the higher-order interactions ($m>2$) are included in a standard mean-field 
approach. In contrast to the fermion gap,
however, the value of the Bertsch parameter appears to be less sensitive to the inclusion
of such higher-order interactions.}
At this point, we would like to remind the reader that we are not aiming at an exact
determination of these quantities in the present work but only at an understanding of the scaling behavior 
of these quantities associated with a variation of~$J$ and~$L$. 
For finite values of~$J$, we have two scales in our theory, namely~$k_{\rm F}$ and~$J$.
The pairing source~$J$ yields a contribution to the density and therefore~$k_{\rm F}$ depends implicitly on~$J$. In 
addition,~$J$ may contribute directly to the fermion gap, i.~e.~$\Delta =\Delta(k_{\rm F},J)$. 
As a consequence, the dimensionless gap~$\Delta/\epsilon_{\rm F}$ is no longer a constant
but depends on the actual values of the source~$J$ and the Fermi momentum~$k_{\rm F}$.
For (small) finite values of~$J$, expanding Eq.~(\ref{eq:xidef}) we find that
the relative shift of Bertsch parameter~$\delta \xi$ takes the form
\be
\!\!\!\delta\xi=\frac{\xi_J\!-\!\xi_0}{\xi_0} =-\frac{2}{3}(\delta n_J)\!+\!\frac{5}{9}(\delta n_J)^2
\!+\!{\mathcal O}\left((\delta n_J)^3\right),\label{eq:deltaxi}
\ee
where~$\xi_0\equiv \xi_{J=0}$ and~$\delta n_J =(n_J - n_{J=0})/n_{J=0}$. Here, $n_{J}$ denotes the density in the 
presence of the source~$J$.  
Recall that the Bertsch parameter is independent of~$k_{\rm F}$ for~$J\to 0$. Thus, the variation 
of~$\delta\xi$ can be expressed solely in terms of the shift of the density~$\delta n_{J}$ which depends on
both~$k_{\rm F}$ and~$J$. We add that~$\delta n_{J}$ tends to zero for 
increasing~$k_{\rm F}$, if we keep the source~$J$ fixed. 
In fact, an increase of~$k_{\rm F}$ for fixed~$J$ corresponds to an increase
of the total density while effectively keeping the contribution from the source~$J$
fixed. On the other hand,~$\delta n_J$ becomes larger for increasing~$J$ and fixed~$k_{\rm F}$. 
\begin{figure}
\includegraphics[width=1\linewidth]{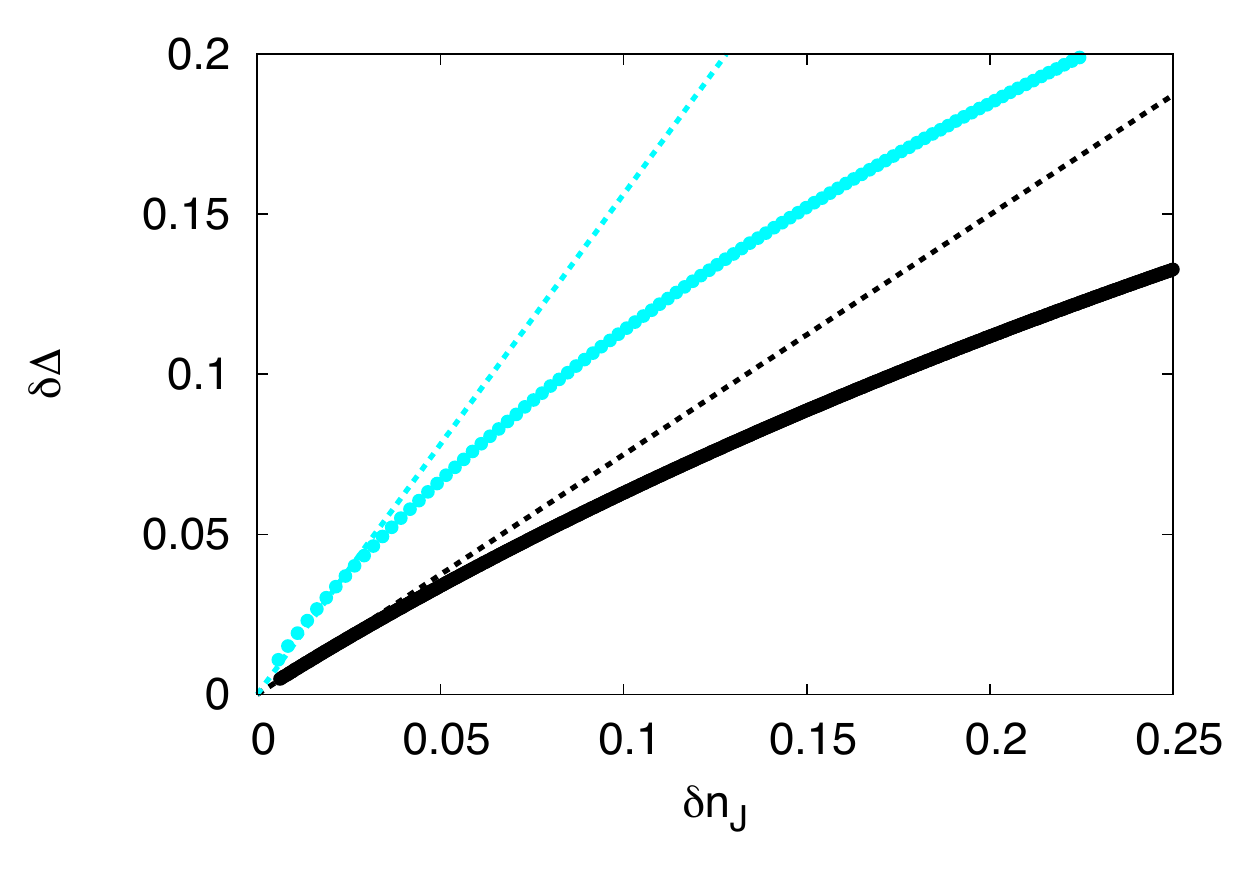}
\caption{\label{fig:dgapdn} Relative shift of the  fermion gap~$\delta\Delta$ 
as a function of~$\delta n_J=(n_J-n_{J=0})/n_{J=0}$ in the thermodynamic limit. The black solid line
depicts the numerical results from our mean-field study for~$J\mu^{7/4}=0.003\dots 0.3$~(or equivalently~$J k_{\rm F}^{7/2}=0.007\dots 0.7$.)
The black dotted line represents the result of the expansion of~$\delta\Delta$ up to first order in~$\delta n_J$, 
where the expansion coefficient has been obtained from a fit to the numerical data. The corresponding results
from an approximation beyond the mean-field limit are depicted by the cyan-colored (light gray) lines.
}
\end{figure}

The expansion coefficients in Eq.~\eqref{eq:deltaxi} are {\it exact}, i.e. the general form of the expansion does not rely on the mean-field approximation. 
In fact, the expansion coefficients are independent of the approximation scheme. 
In principle,~$\delta n_J$ can be also expanded in powers of~$J$.
For an analysis of lattice data, however, an expansion in powers of~$\delta n_J$ seems to be more suitable. 
In Fig.~\ref{fig:dbdn} we show~$\delta \xi$ as a function of~$\delta n_J$ as obtained from our mean-field 
approximation (black/solid line). The linear approximation and the quadratic approximation are given
by the red/dotted line and the blue/dashed line, respectively. We observe that the linear approximation
is justified for~$\delta n_J \lesssim 0.05$. For~$\delta n_J \gtrsim 0.05$, higher order corrections become important.
The knowledge of the range of validity of the linear approximation might be useful for the analysis of 
data from lattice simulations. In these simulations only data for finite values of~$J$ might be available 
such that the limit~$J\to 0$ needs to be extracted from a fit to the available data obtained for~$J\neq 0$. The 
expansion~\eqref{eq:deltaxi} may be well-suited for such a fit since only~$n_{J=0}$ and~$\xi_0$ enter as free (fit)
parameters.

For the dimensionless fermion gap~$\bar{\Delta}=\Delta/\epsilon_{\rm F}$ one might be tempted to 
consider an expansion in powers of the (dimensionless) source~$j=Jk_{\rm F}^{-7/2}$.
Such an expansion seems to be natural for this quantity since 
the source~$J$ contributes directly to the fermion gap~$\Delta$. This can be immediately seen
from the effective action~\eqref{eq:Gmf} and the fact that~$\varphi_1\sim\psi_{\uparrow}\psi_{\downarrow}$. From
a practical point of view, however, it seems to be more appropriate to also consider an expansion
in powers of~$\delta n_J$ rather than~$j$:
\be
\delta\Delta=\frac{\bar{\Delta}_J\!-\!\bar{\Delta}_0}{\bar{\Delta}_0}
=\delta^{(1)}_{\Delta}(\delta n_J) 
+ {\mathcal O}((\delta n_J)^2)\,,\label{eq:deltagap}
\ee
where the~$\delta^{(1)}_{\Delta}$ denotes a dimensionless expansion coefficient.
In Fig.~\ref{fig:dgapdn} we present our results for the relative shift of the gap~$\delta \Delta$
as a function of~$\delta n_J$. From a linear fit to our {mean-field data,\footnote{For the fit, we have used  
25 equidistant values of~$\delta n_J$ between~$\delta n_J\approx 0.006$ and~$\delta n_J\approx 0.015$.}
we find~$\delta^{(1)}_{\Delta}\approx 0.749$, 
see red/dotted line in Fig.~\ref{fig:dgapdn}. A (linear) fit to our results from a study beyond the 
mean-field limit\footnote{Here, we have used  
10 equidistant values of~$\delta n_J$ between~$\delta n_J\approx 0.006$ and~$\delta n_J\approx 0.011$ for the fit.} 
(see Sect.~\ref{sec:bmf}) yields~$\delta^{(1)}_{\Delta}\approx 1.563$.} 
We conclude that the expansion coefficients depend on the truncation scheme. In fact, 
our results suggest that long-range order-parameter fluctuations absent in a mean-field study tend
to increase the slope~$\delta^{(1)}_{\Delta}$.
An exact determination of this coefficient is beyond the scope of this work. We only state
that the linear approximation is in good agreement with the numerical data for~$\delta n_J \lesssim 0.4$
in both cases.
\begin{figure}
\includegraphics[width=1\linewidth]{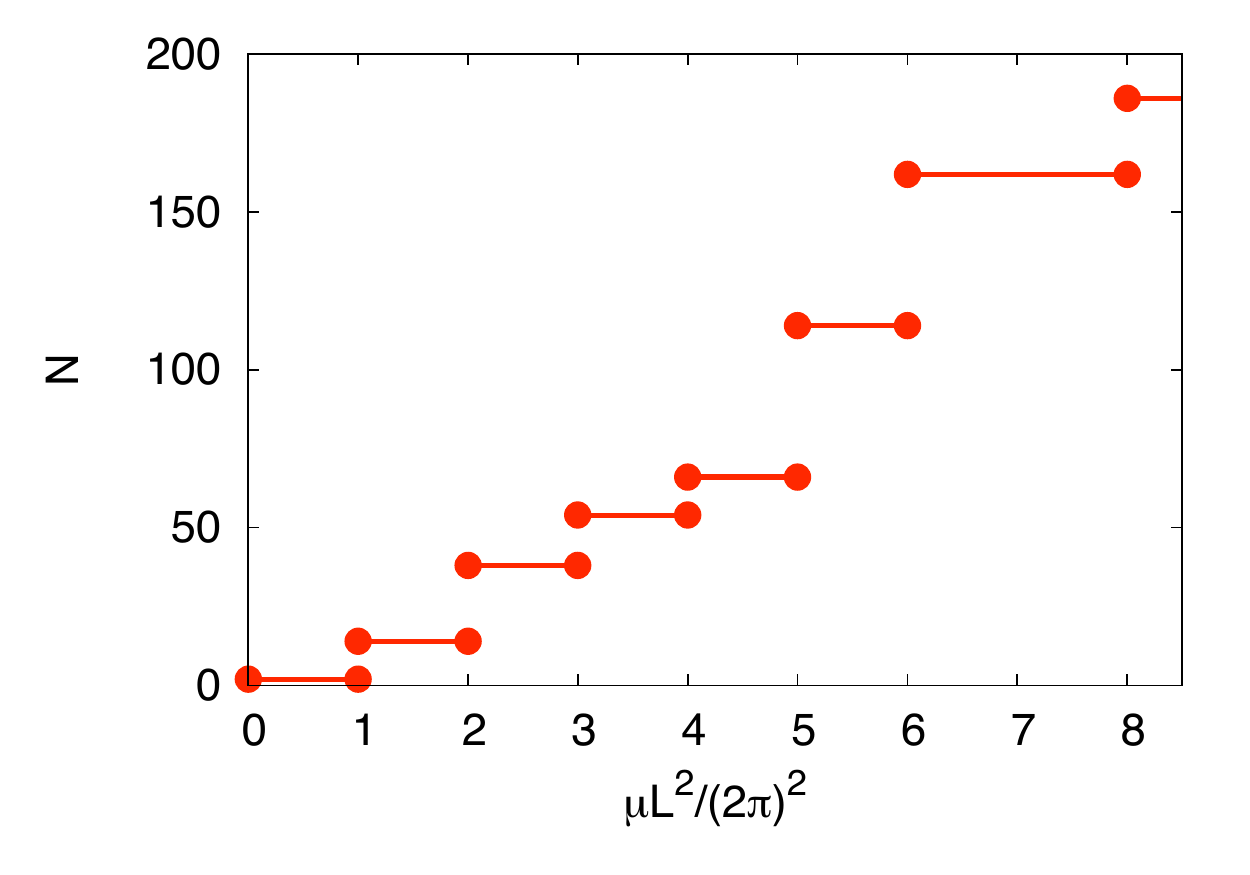}
\caption{\label{fig:nfree} Number of spin-up and spin-down fermions~$N=2N_{\uparrow}=2N_{\downarrow}$ of 
a free Fermi gas as a function of~$\mu L^2/(2\pi)^2$.}
\end{figure}

Now that we have clarified the role of the source~$J$ in the thermodynamic limit,
we turn to a discussion of the effects of a finite volume (in the presence of the source~$J$).
Since we consider a Fermi gas in a volume~$V=L^3$, the momenta of the fermions are discrete. 
These discrete momenta define a 
lattice of equidistant sites in momentum space. A specific momentum then corresponds to 
a specific site of this lattice. 
In Fig.~\ref{fig:nfree} we show the (average)\footnote{For a free Fermi gas at vanishing
temperature,~$N$ only assumes integer values.}
number of particles~$N=n_{\text{free}}L^3$ of a non-interacting gas as 
a function of the dimensionless quantity~$\mu L^2/(2\pi)^2$. 
Increasing~$\mu L^2$, we find that a jump/discontinuity occurs in the particle number~$N$. At
such a discontinuity the number of modes contributing to the partition function increase. In other words, the 
number of modes enclosed by the Fermi sphere with radius~$k_{\rm F}$ changes only 
when~$\mu L^2$ assumes specific values. From Eq.~\eqref{eq:nfree} it follows that~$\mu L^2/(2\pi)^2 \in \mathbb{N}$ is 
a necessary condition for a disconituity in the particle number~$N$.
For~$\mu L^2/(2\pi)^2 \not\in \mathbb{N}$, $N$ is constant since the number of momentum modes
enclosed by the Fermi sphere remains constant. However, we observe that there exists a subset
of integer values for which no jump in the particle number~$N$ occurs, 
e.~g.~$\mu L^2/(2\pi)^2=7$ and~$\mu L^2/(2\pi)^2=15$. For these values of~$\mu L^2/(2\pi)^2$, 
no 3-tuple~$(n_1,n_2,n_3)$ exists such that the sum~$\vec{n}^{\, 2}=n_1^2+n_2^2+n_3^2$ is identical to these values. 
Thus, the particle number remains constant. 
In any case, we refer to a jump in the particle number (density) as a shell effect in the following.
\begin{figure}
\includegraphics[width=1\linewidth]{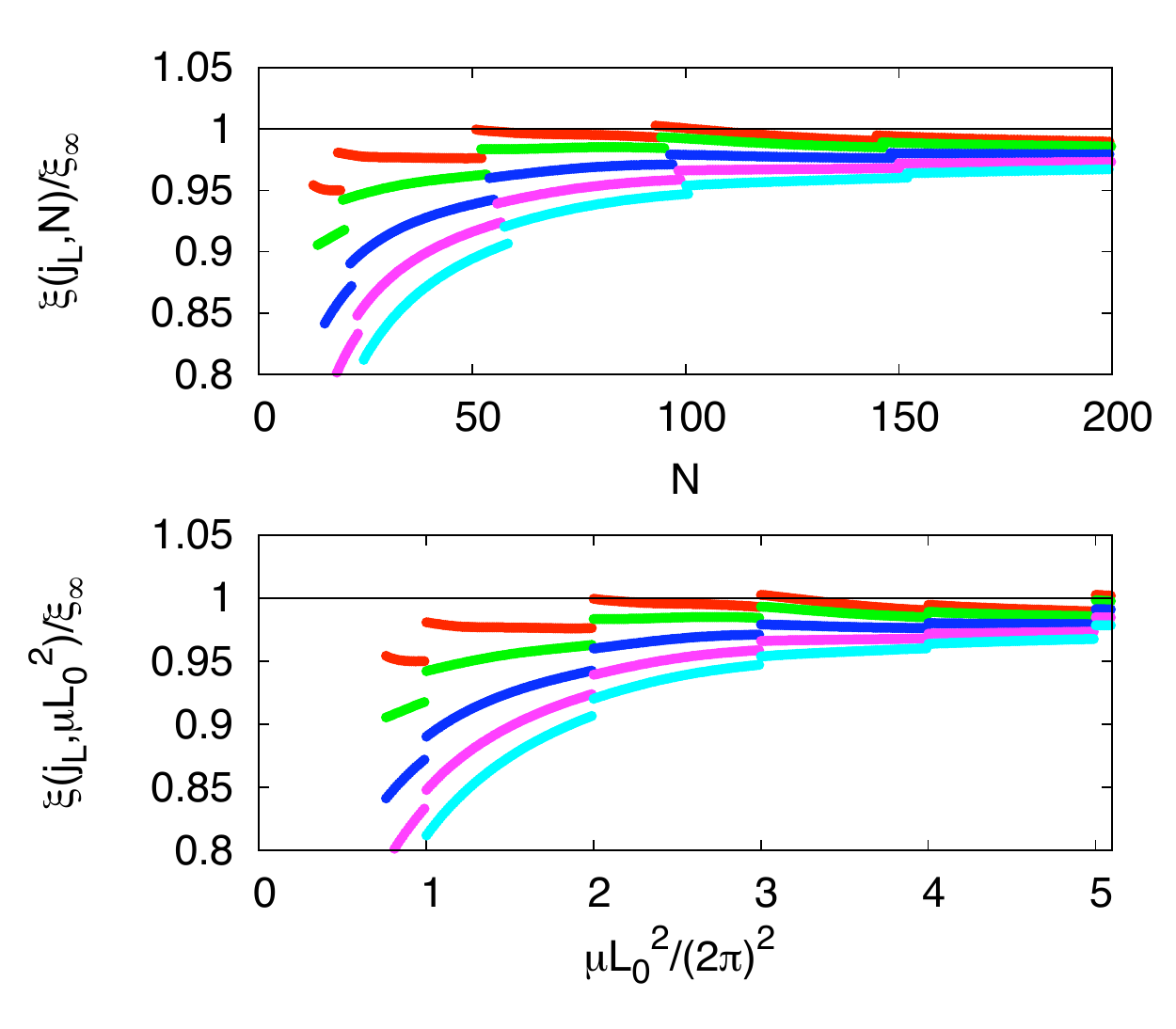}
\caption{\label{fig:mfl10} Top panel: Normalized Bertsch parameter~$\xi (j_L,N)/\xi_{\infty}$ as a function
of the particle number~$N$ for~$j_L=JL_0^{7/2}=0.1,0.25,0.5,0.75,1.0\times 10^{7/2}$ (from top to bottom)
for a fixed spatial extent~$L$ of the cubic volume,~$L=L_0=10\,({\rm c}/{\rm eV})$. Note that different values of~$N$
correspond to theories with different densities.
Bottom panel: Normalized Bertsch parameter~$\xi (j_L,\mu L_0^2)/\xi_{\infty}$ as a function 
of~$\mu L_0^2/(2\pi)^2$ for the same values  of~$JL_0^{7/2}$.
}
\end{figure}

Let us now discuss the dependence of {\it universal} quantities~${\mathcal O}/\epsilon_{\rm F}$ 
on~$J$ and~$L$, where~${\mathcal O}$ has the dimension of energy. 
In this section, we only discuss finite-size and particle-number effects on the Bertsch parameter.
The corresponding effects on the fermion gap will be discussed in the next section. We shall see that the gap
shows a stronger dependence on the volume size, depending on the actual value of~$J$.
In Fig.~\ref{fig:mfl10} we present our results for the (normalized) Bertsch parameter~$\xi/\xi_{\infty}$ as a function
of the (average) particle number~$N$ and as a function~$\mu L^2/(2\pi)^2$ for various values of the 
dimensionless source~$j_L=JL^{7/2}$. Here,~$\xi_{\infty}$ denotes the 
Bertsch parameter for~$J\to 0$ in the thermodynamic limit. In order to obtain these results we have fixed the
spatial extent~$L$ of the volume,~$L=L_0$. In terms of absolute values, we have 
chosen~$L_0=10\,{\rm c}/{\rm eV}\approx 1.97\times 10^{-7}\,{\rm m}\gg a_{\rm B}$, 
where~$a_{\rm B}\approx 5.3\times 10^{-11}\,{\rm m}$ is the Bohr radius, see Ref.~\cite{Diehl:2009ma} for our conventions. 
This choice for~$L_0$ fixes the scale in our study and implies that we measure~$\mu$ as well as~$J$ 
in units determined by the length scale~$L_0$. The density is given by~$n=N/L_0^3$. 

Since we keep~$J$ and~$L$ fixed in Fig.~\ref{fig:mfl10}, our results for the Bertsch parameter depend solely on~$\mu$. Thus, an 
increase of the particle number~$N$ corresponds to an increase in the chemical potential~$\mu$. For large 
values of~$\mu$, we have~$J\mu^{-7/4}\sim (\mu L_0^2)^{-7/4}\to 0$ for a fixed source~$J$ and~$L=L_0$. Thus, we find~$\xi(j_L,N)\to\xi_{\infty}$ 
for~$\mu L_0^2 \gg 1$. For small values of~$N$, shell effects (i.~e. discontinuities) are clearly visible in our results
for the Bertsch parameter. For increasing particle number~$N$ (and fixed~$L_0$), however, we find that 
shell effects are washed out and the 
Bertsch parameter converges rapidly to its value in the thermodynamic limit. 
For the studied values of~$JL_0^{7/2}$ 
and~$N\gtrsim 50$ ($\mu L_0^2/(2\pi)^2 \gtrsim 2$), the relative shift~$\delta\xi$ of the Bertsch parameter
behaves as\footnote{Note that~$k_{\rm F}L\sim N^{1/3}$.}
\be
\delta \xi \sim c_{\rm N}N^{-\frac{2}{3}}\sim c_{\mu}\left(\mu L_0^2\right)^{-1}\,,
\ee
where~$c_{\rm N}<0$ and~$c_{\mu}<0$ are constants. 
The large-$N$ behavior of~$\delta\xi$ follows immediately from the definition of the Bertsch parameter, 
see Eq.~\eqref{eq:xidef}. Moreover, we find 
that~$\xi(j_L,N)/\xi_{\infty}$ is already close to its value in the thermodynamic limit ($\xi(j_L,N)/\xi_{\infty}\gtrsim 0.98$), 
if~$\mu L^2/(2\pi)\gtrsim 5$ ($k_{\rm F}L\gtrsim 18$), i.~e.~$N\gtrsim 200$ for 
the studied values of the source~$J$. In the next section we shall discuss
how order-parameter fluctuations alter these predictions. In any case, the convergence behavior 
depends on~$J$. However, this dependence appears to be weak for the values of~$J$ considered in this work.
\begin{figure}
\includegraphics[width=1\linewidth]{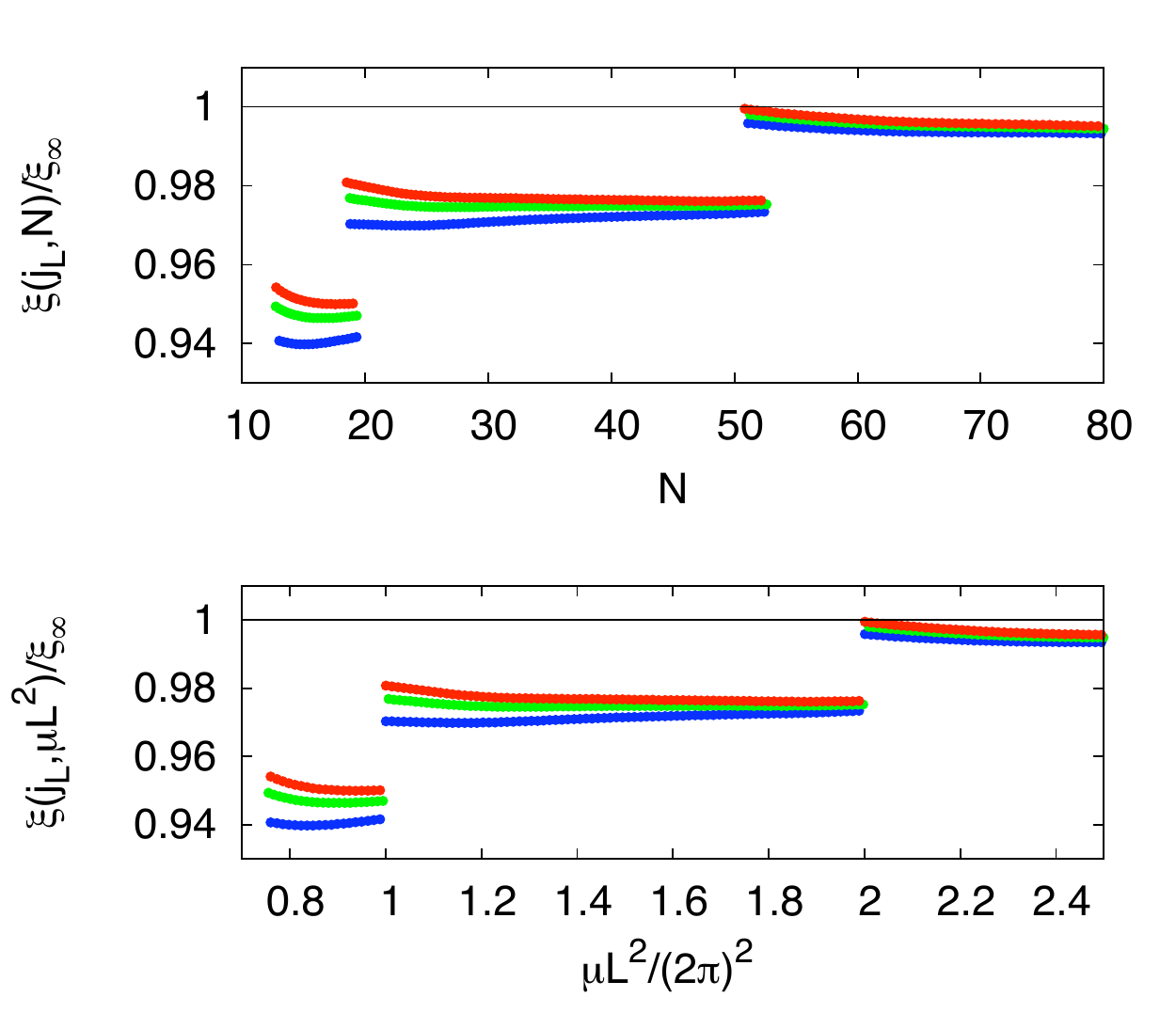}
\caption{\label{fig:mflcomp} Top panel: Normalized Bertsch parameter~$\xi (j_L,N)/\xi_{\infty}$ as a function
of the particle number~$N$ for~$L=5,7.5,10\,\text{c/eV}$ (from bottom to top) for a fixed value 
of the dimensionless source~$j_L=JL^{7/2}=0.1\times 10^{7/2}\approx 316$. 
Bottom panel: Normalized Bertsch parameter~$\xi (j_L,N)/\xi_{\infty}$ as a function
of~$\mu L^2/(2\pi)^2$ for~$L=5,7.5,10\,\text{c/eV}$ (from bottom to top) for a fixed value 
of the dimensionless source~$j_L=JL^{7/2}\approx 316$. 
}
\end{figure}

For a fixed value of~$N < 200$, we observe a clearly visible dependence of~$\xi(j_L,N)/\xi_{\infty}$ on the source~$J$. In
fact,~$\xi(j_L,N)/\xi_{\infty}$ becomes smaller for increasing~$J$. This behavior is compatible with our observations in the thermodynamic
limit where we have also found that~$\xi$ becomes smaller for increasing values of the
source term. For finite values of~$N$, we find that the dependence of the Bertsch parameter with~$J$ 
depends on the actual value of~$\mu L_0^2$. In fact, this dependence becomes stronger for smaller values
of~$\mu L_0^2$. Note that a fixed value of~$\mu L_0^2$ does 
not correspond to a fixed (average) particle number~$N$. This can be readily seen from Fig.~\ref{fig:mfl10}:
In the bottom panel, the actual positions of the discontinuities in~$\xi(j_L,\mu L_0^2)/\xi_{\infty}$ do not
depend on~$J$. On the other hand, the positions of the
discontinuites in~$\xi(j_L,\mu L_0^2)/\xi_{\infty}$ depend on the particle number~$N$ (i.~e. on the density~$n=N/L_0^3$),
see top panel of Fig.~\ref{fig:mfl10}.

In Fig.~\ref{fig:mflcomp} we show~$\xi/\xi_{\infty}$ as a function of~$N$ as well as of~$\mu L^2$ 
for~$L=L_0/2,3L_0/4,L_0$, where the value of the dimensionless source~$j=JL^{7/2}$ has been kept fixed. We find again
that the results become independent of the actual value of~$j$ for large~$N$. For a small fixed value of $\mu L^2$ ($\sim$~small $N$), 
we observe that the Bertsch parameter decreases when the volume size is decreased. This is similar to 
the behavior of~$\xi$ found in the infinite-volume limit, see Eq.~\eqref{eq:deltaxi}. Keeping~$\mu L^2$ 
as well as~$JL^{7/2}$ fixed, we find that the (average) particle number increases when we decrease the 
size of the volume. Since we have $\xi = \mu L^2/(3\pi^2 N)^{2/3}$, it follows that the Bertsch parameter
becomes smaller in smaller volumes, provided that we keep~$\mu L^2$ fixed.
\begin{figure}
\includegraphics[width=1\linewidth]{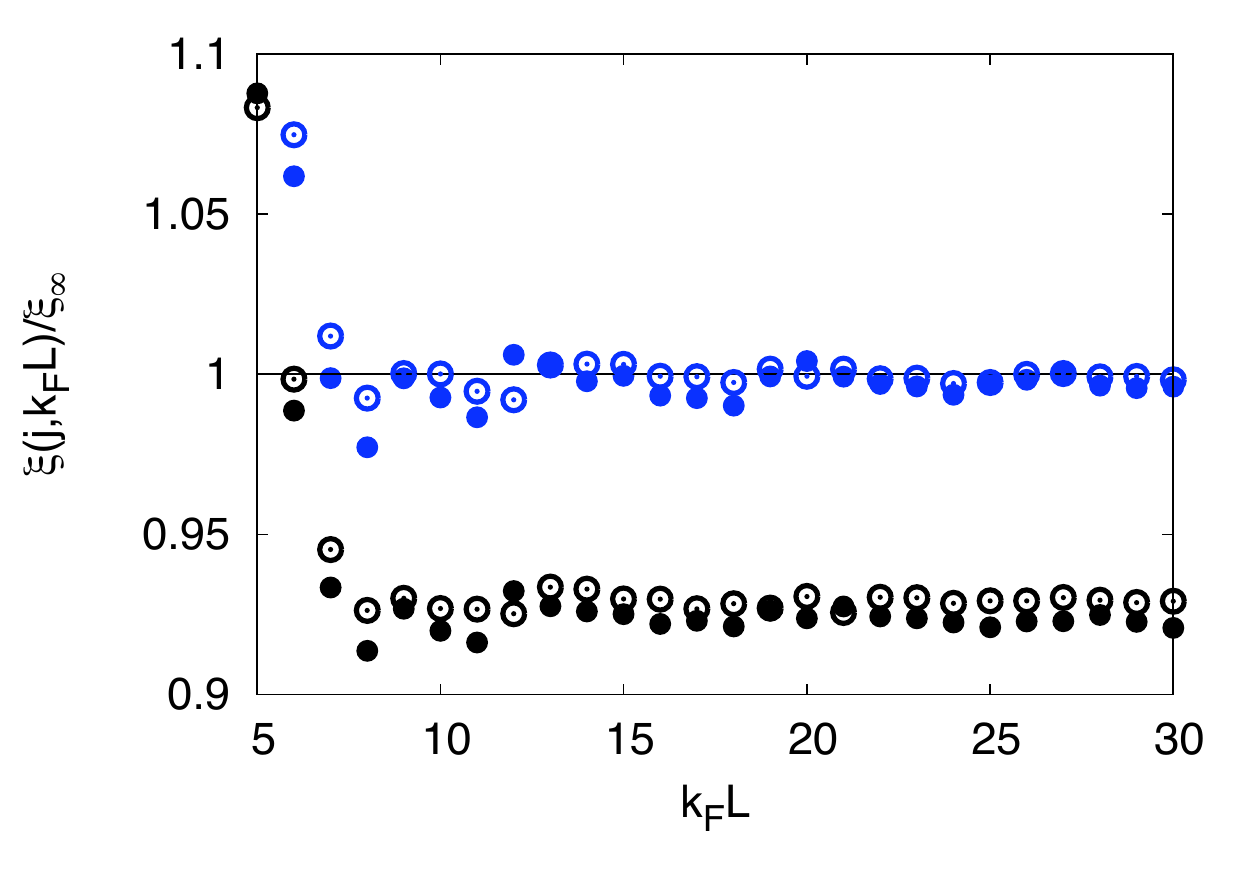}
\caption{\label{fig:b_fixedn} (Normalized) Bertsch parameter as a function of~$k_{\rm F}L$ for a fixed density~$n$. The blue symbols depict
the results for~$j=Jk_{\rm F}^{-7/2}\approx 0.009$ with~$k_{\rm F}=2\,{\rm eV/c}$ ($n\approx 35.2\times 10^{12}/\text{cm}^3$),
whereas the black symbols depict the results for~$j=Jk_{\rm F}^{-7/2}=0.5$ with~$k_{\rm F}=1\,{\rm eV/c}$
($n\approx 4.4\times 10^{12}/\text{cm}^3$).
Open and filled symbols correspond to the mean-field approximation and to our approximation 
including order-parameter fluctuations, respectively. 
}
\end{figure}

Finally, we would like to discuss the behavior of the Bertsch parameter as a function of 
the dimensionless quantity~$k_{\rm F}L$ for a fixed density. In Fig.~\ref{fig:b_fixedn} we show
our corresponding results for~$\xi(j,N)/\xi_{\infty}$ for various values of~$J$, where~$j=Jk_{\rm F}^{-7/2}$.
Filled circles depict our mean-field results.
The results from an approximation beyond the mean-field limit (open circles) 
will be discussed in the next section. For small values
of~$k_{\rm F}L$, finite-size effects are clearly visible. In this regime, the behavior of~$\xi/\xi_{\infty}$
as a function of~$k_{\rm F}L$ is dominated by shell effects. As discussed above, these shell effects are washed
out for large values of the source. For large values of~$k_{\rm F}L$, the Fermi gas approaches the thermodynamic limit,
as it should be for a fixed density. This limit corresponds to a large-$N$ limit since~$k_{\rm F}L\sim N^{1/3}$.
Note the difference to Fig.~\ref{fig:mfl10}, where we have fixed the size of the volume
rather than the density. In the present case, the Bertsch parameter does not approach its value for~$J\to 0$ in 
the thermodynamic limit for large values of~$k_{\rm F}L$. This is due to the fact that~$\xi$ depends on~$J$
for a fixed density, even in the limit~$L\to \infty$, see Fig.~\ref{fig:dbdn}. As in our study 
of~$\xi(j_L,N)/\xi_{\infty}$ as a function of~$N$, we observe that the value of~$k_{\rm F}L$ above which~$\xi$
has effectively assumed its value in the thermodynamic limit depends slightly on the value of the source~$J$. 
In any case, we conclude that finite-size and particle number effects on the Bertsch parameter are rather weak.
\begin{figure}
\includegraphics[width=1\linewidth]{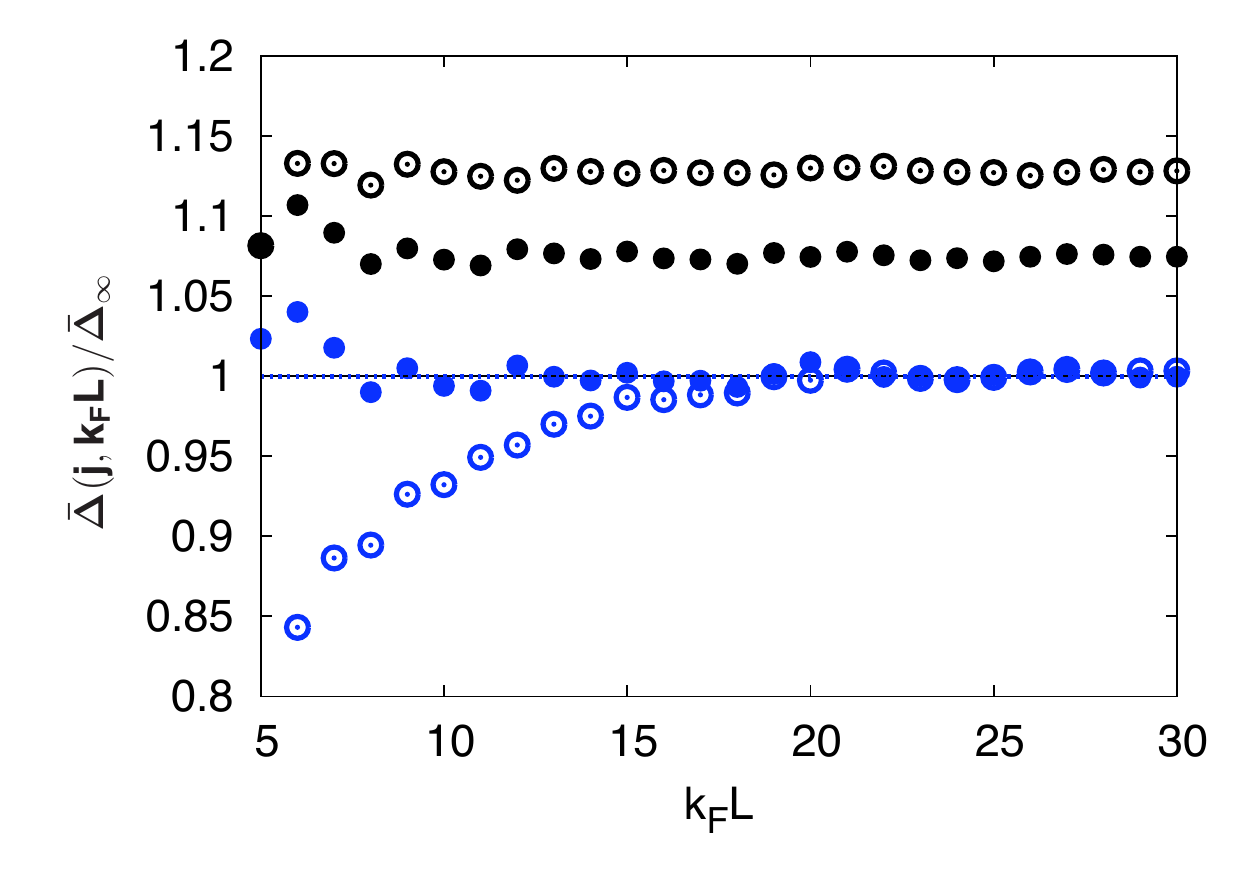}
\caption{\label{fig:gap_fixedn} (Normalized) Dimensionless fermion gap as a function of~$k_{\rm F}L$ for a fixed density~$n$. 
The blue symbols depict the results for~$Jk_{\rm F}^{-7/2}\approx 0.009$ with~$k_{\rm F}=2\,{\rm eV/c}$ ($n\approx 35.2\times 10^{12}/\text{cm}^3$),
whereas the black symbols depict the results 
for~$j=Jk_{\rm F}^{-7/2}=0.5$ with~$k_{\rm F}=1\,{\rm eV/c}$ ($n\approx 4.4\times 10^{12}/\text{cm}^3$). Open and filled symbols
correspond to the mean-field approximation and to our approximation including order-parameter fluctuations, respectively.
{We add that, in the mean-field approximation,}~$Jk_{\rm F}^{-7/2}\approx 0.009$ corresponds
to~$\delta n_J \approx 0.006$ in the infinite-volume limit, whereas~$j=Jk_{\rm F}^{-7/2}=0.5$ corresponds
to~$\delta n_J \approx 0.154$ in the infinite-volume limit.
These different values for~$\delta n_J$ then translate into different values
for the fermion gap in the thermodynamic limit, see Fig.~\ref{fig:dgapdn}. 
}
\end{figure}
\section{Beyond mean-field and the role of fluctuations in a finite volume}\label{sec:bmf}
In this section, we discuss the effects of corrections beyond the mean-field approximation. 
These corrections are associated with fluctuations of the Nambu-Goldstone bosons. The latter 
arise due to the spontaneous breakdown of the~U($1$) symmetry and dominate the IR physics. 
We shall see that such fluctuations play a prominent role in a study of finite-size and particle-number effects.
Whereas~$\langle \phi \rangle_{J=0}$ is not necessarily zero in a finite volume
in the mean-field approximation due to the absence of Nambu-Goldstone fluctuations, we 
expect~$\langle\phi\rangle_{J} \to 0$ for~$J\to 0$ for any finite value of~$L$, see our 
discussion in the previous section. This behavior 
can be indeed observed in our RG analysis once order-parameter fluctuations
are taken into account, cf. also previous RG studies of finite-volume effects in QCD low-energy 
models~\cite{Braun:2004yk,Braun:2005fj,Braun:2005gy,Braun:2010vd}. 

Before we discuss our numerical results in detail, let us briefly comment on the approximation which underlies
our studies in this section. Here, we do not discuss the derivation of the flow equations. Details concerning
the derivation of the flow equations as well as explicit expressions thereof can be 
found in~Refs.~\cite{Diehl:2007XXX,Diehl:2007th,Floerchinger:2008qc,Diehl:2009ma,Scherer:2010sv}. 
In this work, we have simply generalized the flow equations given in~Ref.~\cite{Diehl:2009ma} to the 
case of a finite source~$J$ and a finite volume~$V=L^3$. To be more specific, we include the bosonic loops
in the RG flow of the order-parameter potential. Moreover, we take into account the running of the
wave-function renormalization~$Z_{\varphi}^{\|}$, see our discussion above. Such an approximation
is sufficient for an initial RG study of the effect of order-parameter fluctuations  
on physical observables in a finite volume.
Analogously to our mean-field analysis, finite-volume effects are taken into account by 
replacing the (continuous) squared spatial (loop) momenta~$\vec{q}^{\,2}$ with discrete 
momenta~$\vec{q}^{\,2}=(2\pi)^2\vec{n}^{\,2}/L^2$.
The source~$J$ is again included in the RG flow on all scales~$k$. This 
renders the mass~$m_{\rm G}$ of Nambu-Goldstone
bosons finite and yields an additional contribution to the mass~$m_{\rm R}$ of the radial mode: 
\be
m_{\rm G}^2=\frac{|J|}{\sqrt{2\bar{\rho}_0}}\,,\qquad
m_{\rm R}^2= m_{\rm G}^2 + 2\bar{\lambda}_{\varphi}\bar{\rho}_0\,.\label{eq:massdef}
\ee
As in our mean-field study, the fermion gap has only an implicit dependence on the source~$J$.
The explicit form of the flow equations for the order-parameter potential and the bosonic wave-function
renormalization can be found in App.~\ref{sec:app}.

We begin our discussion of the effect of dynamical Nambu-Goldstone bosons with an analysis of the
(normalized) Bertsch parameter~$\xi(j,k_{\rm F}L)/\xi_{\infty}$ as a function of~$k_{\rm F}L$, 
where~$j=Jk_{\rm F}^{-7/2}$, see~Fig.~\ref{fig:b_fixedn}.
In the thermodynamic limit and for~$J\to 0$, we find~$\xi_{\infty}\approx 0.55$ in the present approximation.
We observe that our results for~$\xi(j,k_{\rm F}L)/\xi_{\infty}$ are in a good agreement with the corresponding 
results from the mean-field approximation. For the presently studied
values of~$J$ and~$k_{\rm F}$ we find that finite-volume effects are clearly visible for~$k_{\rm F}L \lesssim 8$.
For a given value of~$J$ and~$k_{\rm F}L \gtrsim 8$ ($N\gtrsim 20$) the (normalized) Bertsch parameter $\xi(j,k_{\rm F}L)/\xi_{\infty}$ 
then converges rapidly to its value in the thermodynamic limit.

While our present results for the relative shift of the Bertsch parameter in a finite volume 
are in reasonable agreement with our mean-field results, we find that the fermion gap 
is more sensitive to the inclusion of order-parameter fluctuations. In Fig.~\ref{fig:gap_fixedn} we show our 
results for the fermion gap as a function of~$k_{\rm F}L$ for a fixed density. Filled circles depict the 
mean-field results. The results from our study beyond the mean-field limit are depicted by open circles. First of all,
we observe shell effects in the fermion gap as well. However, these effects are less pronounced than in the case of the Bertsch parameter.
This can be understood from the fact that the Bertsch paramter is directly related to the density, while the fermion gap
depends only implicitly on the density of the gap.
In the limit~$k_{\rm F}L\gg 1$ our results for the fermion gap converge to their values in the thermodynamic limit,
as it should be. However, the fermion gap is different for different values of~$J$, see Fig.~\ref{fig:dgapdn}. 
For~$J\to 0$ and~$k_{\rm F}L\to\infty$, we find~$\bar{\Delta}_{\infty}=\Delta/\epsilon_{\rm F}\approx 0.58$ in the present approximation.
For large values of the dimensionless source~$Jk_{\rm F}^{-7/2}$, we observe that the gap is almost independent
of~$k_{\rm F}L$. In this regime, the wave-length associated with the 
lightest excitations in the spectrum, namely the Nambu-Goldstone bosons,
is (much) smaller than the extent~$L$ of the volume, i.~e.~$1/m_{\rm G}< L$. Thus, fluctuations of the Nambu-Goldstone 
particles are not strongly affected by the presence of a boundary. In fact, we observe that corrections beyond the mean-field approximation only 
affect the absolute value of the fermion gap in this regime, whereas the qualitative dependence on~$k_{\rm F}L$ remains unchanged
compared to the mean-field limit. Lowering~$Jk_{\rm F}^{-7/2}$, we find that the fermion gap already starts to deviate from
its value in the thermodynamic limit for large values of~$k_{\rm F}L$. To be specific, we find that finite-volume 
effects are clearly visible for~$k_{\rm F}L \lesssim 15$ ($N\lesssim 110$), see black symbols in Fig.~\ref{fig:gap_fixedn}. 
We add that shrinking the volume size roughly corresponds to an
increase of the temperature in the Euclidean formalism. Therefore the decrease of the gap for small values of~$k_{\rm F}L$ 
can be viewed as a ``melting" of the condensate~$\langle \phi\rangle_{J}$.

Finally, we would like to discuss the approach to the thermodynamic limit for a situation in which the 
limit~$k_{\rm F}L$ might not be easily accessible numerically, e.~g. in lattice simulations.
In principle, the results in Fig.~\ref{fig:gap_fixedn} suggest two
ways to approach the thermodynamic limit. Apart from the direct approach of simplifying increasing~$k_{\rm F}L$ for a fixed
density, a second approach is opened up by studying the Fermi gas with large values 
of the source~$J$, such that~$1/m_{\rm G} \ll L$. The results for the fermion gap are then essentially identical to their values
in the limit~$k_{\rm F}L\to \infty$. From a (linear) fit to the (dimensionless) fermion gaps 
obtained in this way, we can then extract the (dimensionless) fermion gap for~$J\to 0$, see our discussion 
of Fig.~\ref{fig:dgapdn}. 

From our present results together with the observed insensitivity of the mean field approximation on the size of the system,
we see that the finite-size effects are dominated by the bosonic order parameter fluctuations. 
To conclude this section, we therefore comment on the reliability of the truncation of the bosonic sector. 
Here, we have only included leading-order corrections in the derivative expansion. However,
we expect that corrections from higher orders in the derivative expansion only affect our results for the relative
shift of the Bertsch parameter and of the fermion gap on a moderate quantitative level but not 
qualitatively, since standard power counting arguments should be reliable for this system away from criticality. However, we would like to stress that our RG approach is by no means restricted to a low-order derivative expansion
of the effective action. In the context of ultracold gases, the full momentum-dependence of propagators and
vertices has been resolved in, e.~g., Refs.~\cite{Blaizot:2004qa,Floerchinger:2008qc,Schmidt:2011zu}, where
a vertex expansion of the effective action has been studied.
Apart from dropping higher orders in the derivative expansion, we have not considered the possibility of an inhomogeneous 
ground state in our study. This would indeed require the inclusion of higher orders in the derivative expansion. We expect that
corrections from inhomogeneities are subleading, at least for reasonably large values of~$N$. For small particle
numbers~$N$, on the other hand, inhomogeneities in the ground state may play an important role. Despite these shortcomings
of our analysis, we believe that our present study allows us to explore the onset of finite-size and particle number effects
in Fermi gases and therefore our results may help to understand better finite-size effects in lattice (Monte-Carlo) simulations.
\section{Conclusions and Outlook}\label{sec:conc}
In this work we have analyzed finite-size and particle-number effects 
in a Fermi gas confined in a periodic box~$V=L^3$.
We have computed the dependence of the Bertsch parameter and the fermion gap on the average particle number~$N$ and the volume size~$V$.
Moreover, we have studied the dependence of these observables on an external pairing source~$J$ in a finite volume as well as
in the infinite-volume limit. In the present analysis, we have introduced such a source to control symmetry breaking in  a finite volume.
For our studies, we have employed non-perturbative RG techniques which allow us to go systematically
beyond the mean-field approximation by means of a derivative expansion of the effective action. This opens up the possibility to 
study the effects of order-parameter fluctuations on physical observables, such as
the Bertsch parameter and the fermion gap, in a systematic fashion. We find that there are only mild 
corrections to the volume dependence of the Bertsch parameter arising from the inclusion of such effects, whereas the 
behavior of the fermion gap as a function of~$k_{\rm F}L$ is strongly affected by order parameter fluctuations.
To be specific, the Bertsch parameter is already close to its value in the continuum 
for~$k_{\rm F}L\gtrsim 8$ ($N\gtrsim 20$) for a given fixed density. The observed 
rapid convergence of the Bertsch parameter for~$N\gtrsim 20$ is in accordance with recent results from Quantum Monte-Carlo
simulations of a resonantly interacting Fermi gas in a periodic box~\cite{Forbes:2010gt}. 
On the other hand, we find a significant dependence of the fermion gap on~$k_{\rm F}L$ for 
a fixed density, which becomes stronger for small values of the external source. The observed 
``melting" of the fermion gap for decreasing~$k_{\rm F}L$ is associated with strong fluctuations of the Nambu-Goldstone bosons.

Alternatively to a study of the fermion gap~$\Delta_{J} \sim \langle \phi\rangle_{J}$, which plays the role of an order parameter, one might be
interested in a study of~$\langle |\phi|\rangle$ in the absence of a source~$J$. The latter expectation value is not
necessarily zero in a finite volume for~$J=0$.  
Nonetheless, finite-volume effects are also visible in~$\langle |\phi|\rangle$. In fact, it has been found
in, e.~g., lattice studies of relativistic O($2$)~models that~$\langle |\phi|\rangle$ vanishes for small volume sizes in the same way as it 
vanishes for high temperatures, see e.~g. Ref.~\cite{Kogut:2006gt}. Depending on the setup of a given
lattice simulation of an ultracold Fermi gas, a detailed analysis of the onset 
of finite-volume effects in~$\langle |\phi|\rangle$ might therefore be worthwhile as well. A corresponding RG study 
is deferred to future work.

Finally, we add that the observed volume dependence of the fermion gap may affect the determination of the critical temperature in the
unitary limit in Monte-Carlo simulations. The present work aims to set the stage of a detailed RG scaling analysis of the critical behavior
of a resonantly interacting Fermi gas, including
a study of the dependence of the critical temperature on the volume size and the particle number. Another interesting 
extension of the present RG approach represents a study of a Fermi gas based on the canonical ensemble rather than the
grand canonical ensemble which we have considered here. This might be possible in a similar way as done in lattice QCD simulations,
see e.~g. Ref.~\cite{Kratochvila:2005mk}. Such an analysis may then provide additional insights into the many-body physics of ultracold 
atomic gases and may help to further bridge the gap between Monte-Carlo simulations and continuum approaches.

\emph{Acknowledgment.}  The authors are very grateful to J.~E.~Drut and H.~Gies for critical comments on the manuscript.
Moreover, the authors gratefully acknowledge useful discussions with 
J.~E.~Drut, S.~Floerchinger, H.~Gies, B.~Klein, J.~M.~Pawlowski and A.~Schwenk. 
JB acknowledges financial support by the DFG under Grant BR 4005/2-1 and the DFG research 
training group GRK~1523/1. Moreover, JB and MMS acknowledge support by the DFG research
group {\it Functional renormalization group in correlated systems} (FOR 723).
SD acknowledges support by the Austrian Science Fund (FOQUS), the European Commission (AQUTE, NAMEQUAM), 
and by a grant from the US Army Research Office with funding from the DARPA OLE program.

\appendix

\section{RG Equations Beyond the Mean-Field Limit}\label{sec:app}
Beyond the mean-field approximation, the RG flow of the order-parameter potential~$U$
receives contributions from a purely bosonic loop. To be specific, the flow equation
reads
\be
\partial_t U(\rho,J,L,\mu)&=&\eta_{\varphi}\rho\, U^{\prime}
-2k^5 (B^{>}_{\rm F}+B^{<}_{\rm F})s_{\mathrm{F}} \nn\\
&& \quad+\, k^5 \left(1-\frac{2\eta_{\varphi}}{5}\right)B_{\rm B} s_{\rm B}\,,
\label{eq:mfflowUBMF}
\ee
where~$\rho=Z_{\varphi}^{\|}\bar{\rho}$,~$\eta_{\varphi}=-\partial_t \ln Z_{\varphi}^{\|}$ and
\be
B_{\rm B}\! =\! \frac{1}{(kL)^3}\sum_{\vec{q}}\theta \left(2(kL)^2\!-\! (2\pi)^2\vec{q}^{\,2}\right)
\stackrel{(kL)\to\infty}{\longrightarrow} \frac{\sqrt{2}}{3\pi^2}\,.\nn
\ee
This function counts the (bosonic) momentum modes and corresponds to the mode-counting
functions for the fermions defined in Eqs.~\eqref{eq:BFg} and~\eqref{eq:BFs}. 
Recall that we use periodic boundary conditions for the fields. The function~$s_{\rm B}$ is given by
\be
s_{\rm B}=\frac{1}{2}\left[
\sqrt{\frac{1+\omega_1}{1 + \omega_2}}
+ \sqrt{\frac{1+\omega_2}{1 + \omega_1}}
\,\right]\,,
\ee
where~$\omega_1 = U^{\prime}/k^2$ and~$\omega_2 = (U^{\prime} + 2\rho U^{\prime\prime})/k^2$; 
the prime denotes the derivative with respect to~$\rho$. The quantities~$\omega_1$ and~$\omega_2$ 
are related to the (bare) masses~$m_{\rm G}$ and~$m_{\rm R}$ which we have defined in 
Eq.~\eqref{eq:massdef}:
\be
\omega_1\big|_{\rho_0}=\frac{m_{\rm G}^2}{Z_{\varphi}^{\|}k^2}\,,\qquad
\omega_2\big|_{\rho_0}=\frac{m_{\rm R}^2}{Z_{\varphi}^{\|}k^2}\,.
\ee
Here, we have used that
\be
\bar{m}^2_{\varphi}\langle \phi_1\rangle = J\,,
\ee
see Ref.~\cite{Braun:2010vd}. The RG running of~$Z_{\varphi}^{\|}$ is determined by the anomalous dimension~$\eta_{\varphi}$:
\be
\eta_{\varphi}=\frac{h_{\varphi}^2}{12\pi^2(1+\omega_3)^{\frac{5}{2}}}
\left(B^{>}_{\rm F}-B^{<}_{\rm F}\right) (2-\omega_3)
\,,
\ee
where~$\omega_3= h_{\varphi}^2\rho_0/k^4$. The functions~$B^{>}_{\rm F}$ and~$B^{<}_{\rm F}$ are defined
in Eqs.~\eqref{eq:BFg} and~\eqref{eq:BFs}, respectively.
Note that a non-trivial RG running of~$Z_{\varphi}^{\|}$ induces a non-trivial RG-scale dependence of
the (dimensionless) Yukawa coupling~$h_{\varphi}^2=\bar{h}_{\varphi}^2/(Z_{\varphi}^{\|}k)$, see our discussion in Sect.~\ref{sec:mf}. 

\end{document}